\documentstyle[preprint,aps,eqsecnum]{revtex}

\begin{document}
{\tighten
\preprint{\vbox{\hbox{UCSD/PTH 94--15}}}

\title{Confining Flux Tubes in a Current Algebra Approach}

\author{Katherine M.~Benson and Maha Saadi}
\address{Department of Physics\\
University of California,~San Diego\\  La Jolla, California 92093}

\bigskip
\date{revised October 31, 1994}

\maketitle
\begin{abstract}

  We describe flux tubes and their interactions in a low energy sigma
  model induced by $SU({N_f}) \rightarrow SO({N_f})$ flavor symmetry
  breaking in $SO(N_c)$ QCD.  Unlike standard QCD, this model allows
  gauge confinement to manifest itself in the low energy theory, which
  has unscreened spinor color sources and global $Z_2$ flux tubes.  We
  construct the flux tubes and show how they mediate the confinement
  of spinor sources.  We further examine the flux tubes' quantum
  stability, spectrum and interactions.  We find that flux tubes are
  Alice strings, despite ambiguities in defining parallel transport.
  Furthermore, twisted loops of flux tube support skyrmion number,
  just as gauged Alice strings form loops that support monopole
  charge. This model, while phenomenologically nonviable, thus affords
  a perspective on both the dynamics of confinement and on subtleties
  which arise for global Alice strings.

\end{abstract}

}
\narrowtext

\section{Introduction and Model}

Nonlinear sigma models have had great success in describing low energy
QCD phenomenology.  However, they have not captured the hallmark
feature of QCD: confinement, where the potential between $q\bar{q}$
pairs grows linearly with separation.  Such linear potentials can
arise as tension-carrying flux tubes --- between $q$ and $\bar{q}$,
for Yang-Mills QCD, or between any unscreened sources which persist in
low energy effective theories. The conventional Skyrme model, induced
by the global flavor symmetry breaking $SU({N_f})_L \times SU({N_f})_R
\rightarrow SU({N_f})_{diag}$, supports neither unscreened sources nor
flux tubes. Global flux tubes are classified by $\pi_2 (G/H)$, due to
their constant vev at spatial infinity. This necessarily vanishes
whenever the vacuum manifold $G/H$ is itself a Lie group, as occurs in
the Skyrme model.

Witten \cite{wittenon} noted that an $SO(N_c)$ gauge theory of QCD
induces a different Skyrme model, whose topology {\em can} support
flux tubes. This theory has ${N_f}$ lefthanded quarks $q_L$, which
transform as (real) fundamentals under color $SO(N_c)$.  $SO(N_c)$ has
no gauge anomaly, and thus requires no right-handed quarks; however,
its $U(1)_B$ anomaly breaks baryon number to $Z_2$.  The high energy
theory thus has the full symmetry group $Z_2 \times SO(N_c) \times
SU({N_f})$. At low energies, flavor $SU({N_f})$ breaks to its $SO({N_f})$
subgroup, due to formation of a quark condensate $\langle
q_{Li}^\alpha \,\, \Sigma_{ij} \,\, q_{Lj}^\alpha\rangle$ (where
$\alpha$ and $i$ are color and flavor indices respectively). This
condensate interacts with quark excitations in the theory, inducing
an effective Majorana mass for the quarks
\begin{equation}
\label{LMaj}
{\cal L}_m =         - \mu \left( {\overline{q_L}}  \Sigma  q_{Lc}
         +             {\overline{q_{Lc}}} {\Sigma}^\dagger  q_{L} \right)\ ,
\end{equation}
which displays the $SU({N_f})\rightarrow SO({N_f})$ flavor symmetry breaking
explicitly.  The Goldstone modes $\Sigma$ are described by a
$SU({N_f}) / SO({N_f})$ nonlinear sigma model, with skyrmions
($\pi_3(G/H)$ = $Z_2$, for ${N_f}
\ge 4$) and flux tubes ($\pi_2(G/H)$ = $Z_2$, for ${N_f}
\ge 3$). This defect classification makes physical sense: for baryons,
identified with antibaryons because real quarks are identified with
antiquarks; and for flux tubes, whose $Z_2$ structure emerges as a
response to external spinor sources, which can be screened by
fundamental quarks only in even combinations.

In this paper, we elaborate on our recent work with Manohar
\cite{shortskyrme}, constructing the flux tubes in this theory and showing
that their interactions with skyrmions and spinor sources obey
heuristic expectations.  Confinement of spinor sources in an $SO(N_c)$
gauge theory can thus manifest itself in the low energy sigma model,
through relic phenomenology in the presence of unscreened sources.

We organize our results as follows: section II derives the unique flux
tube form with minimal energy; section III examines its classical
stability and dynamics; section IV its quantum stability and spectrum;
and section V its interactions and relationships with other
fundamental objects in the theory. The early sections reveal that
minimal flux tubes lie in a planar subspace of the vacuum manifold
$SU({N_f})/ SO({N_f})$; while section V discusses why our flux tubes are
Alice strings, despite ambiguities in defining parallel transport
around them; why they carry skyrmion number when twisted; and how
they can be viewed as mediating the confinement of spinor sources.

\section{Finding Nontrivial Flux Tubes}

To construct topological defects, we seek configurations where the
condensate $\Sigma$ varies spatially in both nonsingular and
nontrivial ways. Nonsingular variations assume a form dictated by the
transformation properties of $\Sigma$. Demonstrating their
nontriviality, however, is complicated in our modified Skyrme model.  The
difficulty stems from the same source as the flux tubes themselves:
the fact that the global vacuum manifold $G/H$ is not itself a Lie
group, but only a quotient space.  Thus, unlike skyrmions in the
standard Skyrme model, winding numbers for $\Sigma$ cannot be
obtained from a group structure on $G/H$ alone, as indices
dependent only on $\Sigma$. Instead they must be determined by
homotopy arguments from the embedding of $\Sigma$ in $G$.

The $SU(2)_L \times SU(2)_R$ Skyrme model presents a simpler case.
Here symmetry group transformations $(g_L, g_R)$ act on $G$ by left
multiplication and leave $\Sigma$ invariant under the diagonal
subgroup.  $\Sigma$ can thus be constructed from underlying group
transformations: $\Sigma = g_L {g_R}^\dagger,$ transforming as $\Sigma
\rightarrow L \Sigma {R}^\dagger$ under the group element $(L,R)$. When
$\Sigma = {{\mbox{1\hskip-0.22em\relax l}}\,} ,$ $L=R$ gives the diagonal
unbroken subgroup. For
$\Sigma \ne {{\mbox{1\hskip-0.22em\relax l}}\,},$ however, the embedding $H
\subset G$ changes and
$L=R=g$ rotates $\Sigma$ on the vacuum manifold, $\Sigma \rightarrow g
\Sigma {g}^\dagger.$ Using these transformation laws, and imposing
spherical symmetry to minimize energy, we may write an arbitrary
$\pi_3$ defect as $\Sigma = L(\Omega) \Sigma_o(r) {R}^\dagger
(\Omega)$. We choose our basis to diagonalize $\Sigma_o (r) = \exp
{\{iF(r)\tau_z\} }$, with $F(r \rightarrow \infty) = 0.$ $\Sigma$ then has
finite energy only when $L=R,$ keeping $\Sigma$ constant at spatial
infinity, and is singular at the origin unless $F(0)= n\pi$.

Establishing this nonsingular form $\Sigma = g(\Omega) \exp
{\{i F(r)\tau_z\}} {g}^\dagger (\Omega)$ requires only the quotient
space structure $G/H = \left( SU(2)_L \times SU(2)_R \right)/
SU(2)_{diag}.$ We require the full Lie group structure of $G/H$,
however, to show that a particular $g(\Omega)$ induces nontrivial
$\pi_3$, or equivalently, to reduce the Pontryagin index to a function
of $\Sigma.$ Since $\pi_3(G/H) = \pi_3(SU(2)) = Z,$ the homotopy
index $\pi_3$ simply measures the number of times $\Sigma(r,\Omega )$
covers $SU(2).$ This is easily determined by group integration or
inspection. For the skyrmion, $g(\Omega) = \exp{\{-i\theta \tau_z/2\}}
\exp{\{-i\phi\tau_y/2\}}$ rotates $\tau_z$ to an
arbitrary Lie algebra element $\hat{r}\cdot\hat{\tau}.$ Thus $\Sigma =
\exp{\{i F(r)\hat{r}\cdot\hat{\tau} \}}$ consists of the exponentiation of
the full Lie algebra from $0$ to $n\pi$ --- by definition, covering the
Lie group $n$ times, for a $\pi_3$ winding $n$.

Our model yields a nonsingular ansatz for $\Sigma$ as directly as the
standard one. $G = SU({N_f}),$ acting on itself by left multiplication,
leaves $\Sigma$ invariant under the orthogonal subgroup. We can thus
write $\Sigma$ as $g {g}^T,$ transforming as $\Sigma \rightarrow a
\Sigma{a}^T$ under the group element $a$. When $\Sigma =
{{\mbox{1\hskip-0.22em\relax l}}\,} ,
a \in SO({N_f})$ gives the unbroken subgroup $H$; elsewhere, the
embedding $ H\subset G$ is parallel transported to $g H {g}^\dagger$.
Again, a cylindrically symmetric defect can be written as an
$r$-dependent vev, with angle-dependent group rotation: $\Sigma =
g(\theta) \Sigma_o(r){g}^T (\theta).$ Choosing $\Sigma_o =
{{\mbox{1\hskip-0.22em\relax l}}\,} $ at spatial infinity restricts $g(\theta)$
to lie in the unbroken
$SO(n)$ subgroup, and to commute with $\Sigma_o(r=0),$ if $\Sigma$ is
to be nonsingular with finite energy.

We can further characterize $\Sigma $ in terms of $su({N_f})$ basis
generators. These specify rotations in all 2-dimensional subplanes
$(jk)$, and are usually taken as follows (with Cartan norm ${\mbox{tr}\,} T_a
T_b = {{\textstyle\frac{1}{2}}} \delta_{ab}$). The rank ${N_f}-1$ Cartan
subalgebra has as
basis diagonal matrices $T_d$,
\begin{equation}
\label{defTd}
T_d = ( 2d(d +1)\, )^{-1/2}\ {\rm diag} (1,\ldots,1, -d,0,
\ldots,0),
\end{equation}
with ones in the first $d$ entries and $d = 1$ to ${N_f} -1$.  We can
overspecify this basis, sacrificing the Cartan norm, by the set $\{
{{\textstyle\frac{1}{2}}}\tau_{z\,(jk)\,}\}$ in all subplanes of the ${N_f}$
dimensional
vector space.  Off-diagonal generators $\{
{{\textstyle\frac{1}{2}}}\tau_{x\,(jk)\,},
{{\textstyle\frac{1}{2}}}\tau_{y\,(jk)\,}\}$ complete the $su({N_f})$ basis.
Flavor
symmetry breaking divides the basis into two sets: antisymmetric
matrices $\{T_h\} \equiv\{{{\textstyle\frac{1}{2}}} \tau_{y\,(jk)\,}\}$,
generating $H =
SO({N_f});$ and symmetric matrices $\{T_b\} \equiv\{
{{\textstyle\frac{1}{2}}}\tau_{x\,(jk)\,},
{{\textstyle\frac{1}{2}}}\tau_{z\,(jk)\,} \}$, generating no
symmetries.  $\Sigma_o(r),$ a unitary symmetric matrix, can be written
$\exp {\{i F_b(r) T_b\} }.$ Thus $\Sigma$ assumes the most general
nonsingular form $\Sigma = h(\theta) \exp {\{ i F_b(r) T_b)\} }
h^{-1}(\theta),$ for $h(\theta) \in H$ and $F_b(r)$ ranging from zero
at infinity to $ 2\pi n \,\,\delta_{bb'}$ (for some fixed direction
$b\, '$) at the origin.

Having obtained nonsingular configurations $\Sigma$, we must
demonstrate their nontriviality. Unlike the conventional Skyrme
model, this requires an understanding of how $\Sigma = g
{g}^T$ arises from the underlying group mapping $g$.
For flux tubes, we must construct $\Sigma$ from the exact sequence
\begin{eqnarray*}
\pi_2\left( SU({N_f})\right) = 0  \quad \rightarrow \quad&&
\pi_2\left( SU({N_f})/SO({N_f})\right) \\
&&\quad\quad \rightarrow \quad
\pi_1\left( SO({N_f})\right) = Z_2 \quad \rightarrow \quad
\pi_1\left( SU({N_f})\right) = 0 \quad .
\end{eqnarray*}
That is, $g g^T$ gives a nontrivial $\Sigma$ only if $g$ corresponds
to some mapping from the plane to $SU({N_f}),$ with boundary values in
the $SO({N_f})$ subgroup. Furthermore, when parametrized as a family of loops,
$g$ must start at the identity and end on a nontrivial loop in $SO({N_f})$.
Taking $\left (\alpha\in[0,2\pi], \beta\in[0,\pi]\right) $ as our
coordinates on the plane, these criteria become
\begin{equation}
\label{gbc}
g(\alpha, \beta)  = \left\{
\begin{array}{ll}
{{\mbox{1\hskip-0.22em\relax l}}\,} & \mbox{when $\alpha = 0$, $\alpha = 2\pi$,
\ or $\beta = 0$}\\
h^2(\alpha) & \mbox{when $\beta = \pi$},
\end{array}
\right.  \end{equation}
where $h^2(\alpha)$ is a nontrivial loop in $SO({N_f}).$ Such nontrivial
loops can be written $h^2(\alpha) = \exp {\{ i \alpha \, 2n_h T_h\}
  },$ where $T_h$ is the set $\{{{\textstyle\frac{1}{2}}} \tau_{y(jk)} \}$
introduced
above and $n_h T_h$ generates rotations in a single plane. (Of
course, these loops can be deformed, but deformations from
geodesic form lengthen the loop and induce additional gradient
energy in $\Sigma.$ We discard them, to focus on mappings that
produce minimal gradient energy.)

We now show that the embeddings $g(\alpha, \beta)$ have minimal
gradient energy only when they induce flux tubes $\Sigma$ of a unique
form. We show this by imposing consistency and minimal energy
conditions on the most general trivialization $g(\alpha,
\beta)$. We then relate the topological coordinates $(\alpha, \beta)$ to
cylindrical coordinates $(r,\theta)$, to obtain the physical flux tube
$\Sigma(r,\theta)$. Finally we consider the low-lying deformations of
$\Sigma(r,\theta)$ which can be favored by potential energy terms. We
thus obtain a family of non-trivial flux tubes, among whom a minimal
representative is selected dynamically.

We construct the most general trivialization $g(\alpha,\beta)$ as follows.
Start, at $\beta = 0,$ by left multiplying $h(\alpha)$ by its inverse:
$g( \beta = 0) = h^{-1}(\alpha)\, h(\alpha) ={{\mbox{1\hskip-0.22em\relax
l}}\,}.$ As $\beta$ varies,
allow the left multiplier $h^{-1}(\alpha)$ to vary over the full group
$G$: $g(\alpha, \beta) = g_1(\beta)\, h^{-1}(\alpha) \, g_2(\beta) \,
h(\alpha) $.  The $\alpha=0$ condition and uniqueness of the inverse
imply $g_2(\beta) = g_1^{-1}(\beta).$ To probe structure on $G/H$
(where $a\rightarrow a a^T$ identifies cosets $\{gH\}$),  write
$g_1(\beta)= b(\beta)\,\tilde{h}(\beta),$ for $\tilde{h}(\beta)\in H$ and
$b$ generated by broken generators $T_b$. (Specifically, since
$g_1(\beta) \, g_1^T(\beta) =\exp {\{ i G_b(\beta)T_b\}}$, we can write
$g_1(\beta) = \exp {\{ i G_b(\beta)T_b/2)\}}\,\,\tilde{h}(\beta)$.) Thus
we may write the most general trivialization of the loop $h^2(\alpha)$
in $G$:
\begin{equation}
\label{gformgen}
g(\alpha, \beta) =
b(\beta) \ \tilde{h}(\beta) \,\,h^{-1}(\alpha)\,\, \tilde{h}^{-1}(\beta)
                        \ b^{-1}(\beta) \ h(\alpha).
\end{equation}
This induces minimal energy only in geodesic form
\begin{eqnarray}
\label{gformgeod}
b(\beta) &=& \exp{\{i l \beta \,\, n_b T_b\}} \nonumber\\
\tilde{h}(\beta)  &=& \exp{\{i m \beta\,\, \tilde{n}_h T_h\}} \nonumber\\
h(\alpha)  &=& \exp{\{i \alpha \,\, n_h T_h\}} ,
\end{eqnarray}
where $n_b$ and $n_h,\tilde{n}_h$ are unit vectors over the range of
${T_b}, {T_h}$ respectively.

We now show that consistency reduces the distinct choices of
$b,h,$ and $\tilde{h}$.  For simplicity, take $\tilde{h} =
{{\mbox{1\hskip-0.22em\relax l}}\,}$ at
first.  Choose a basis in $su({N_f})$ so that $n_b T_b =
{{\textstyle\frac{1}{2}}}\tau_{z\,(12)\,}$.  Then, under conjugation by
$b(\beta)$, $2n_h T_h =
 n_{(jk)}\,\, \tau_{y(jk)}$ rotates as follows:
\begin{eqnarray}
\label{rotTh}
b(\beta)\ 2 n_h T_h\ b^{-1}(\beta)&=&
n_{(12)}\ \left(\cos l\beta \,\tau_{y(12)} + \sin l\beta \,\tau_{x(12)}\right)
+
{\sum_{2<j<k}} n_{(jk)}\,\,\tau_{y(jk)}\nonumber\\
&&+ {\sum_{\stackrel{j=1,2}{k>2}}} \, n_{(jk)} \
\left(\cos(l\beta/2) \,\tau_{y(jk)}
- (-1)^j \, \sin(l\beta/2) \,\tau_{x(jk)}\right)  .
\end{eqnarray}
To obtain $g(\alpha,\beta=\pi) = h^2(\alpha)$, this conjugated
generator must give $-2 n_h T_h$ at $\beta = \pi$. This can occur for
only two lowest winding possibilities: $l$ can be 1, with $n_{(jk)} =
e_{(12)}$ vanishing outside the $(12)$ plane; or $l$ can be $2$, with
$n_{(jk)}$ describing a plane that intersects $(12)$ in a single line.
The second possibility is further constrained by the boundary
condition at $\alpha =2\pi$, which requires $h(2\pi)$ to commute
with $n_b T_b$. This occurs only when ${\hbox{\bf [} n_b T_b, (n_h T_h)^2
\hbox{\bf ]} }$
vanishes --- that is, when the plane $n_{(jk)}$ intersects $(12)$
along a coordinate axis.  Thus only three distinct candidates arise
for the pair $b,h$: $l=1, n_{(ij)} = e_{(12)}$; $l=2, n_{(ij)} =
e_{(13)}$; and $l=2, n_{(ij)} = e_{(23)}$.

Allowing nontrivial $\tilde{h}(\beta)$ produces no additional flux tubes
of minimal energy, as we demonstrate explicitly in an appendix.
Thus the form \ref{gformgen} for the flux
tubes's embedding reduces to two candidates:
\begin{equation}
\label{gform}
g(\alpha, \beta) =
b(\beta) \ h^{-1}(\alpha)
                        \ b^{-1}(\beta) \ h(\alpha),
\end{equation}
with
\begin{equation}
\label{defbh}
h(\alpha)  = \exp{\{i \alpha \,\, T_{h\Box}\}}  \quad
b(\beta) =
\left\{ \begin{array}{c}
 \exp{\{i \beta \,\, T_{b\Box}\}}\\
 \exp{\{i 2\beta \,\, T_{b\Box '}\}}
 \end{array}\right.,
 \end{equation}
where $\Box$ and $\Box '$ denote planes intersecting along a
coordinate axis, and $T_b, T_h$ are ${{\textstyle\frac{1}{2}}}\tau_z ,
{{\textstyle\frac{1}{2}}}\tau_y$ in
the indicated planes. Written in this form, we see that the two
candidates are in fact the same. From equation \ref{rotTh}, both
choices for $b(\beta)$ induce the same rotated group element $b(\beta)
\ h^{-1}(\alpha)\ b^{-1}(\beta)$ (modulo sign redefinition of $\beta$
).  We can thus choose a single deformation $g(\alpha, \beta)$,
exploring only a planar $SU(2)$ subgroup of $SU({N_f})$, to produce the
flux tube of minimal gradient energy associated with the $Z_2$ loop
$h^2(\alpha)$. Note that $g$ indeed gives a $Z_2$ object, with
$g(\alpha, \beta = 2\pi) = {{\mbox{1\hskip-0.22em\relax l}}\,}$.  (This $Z_2$
structure appears in the
$\beta \rightarrow -\beta$ equivalences above, as well).

 From this form we construct the flux tube
$\Sigma = g g^T:$
\begin{equation}
\label{Sigbh}
\Sigma = b(\beta) \ h^{-1}(\alpha)
                        \ b^{-2}(\beta) \ h(\alpha) \ b(\beta).
\end{equation}
This gives
\begin{equation}
\label{Sigtpexp}
\Sigma(\alpha, \beta) =
{{\mbox{1\hskip-0.22em\relax l}}\,} - 2 \sin^2{(\alpha /2)} \sin^2{\beta}\
{{\mbox{1\hskip-0.22em\relax l}}\,}_\Box
+  i \sin^2{(\alpha /2)} \sin{(2\beta)}\ \tau_{z\Box}
-  i \sin{\alpha } \sin{\beta}\ \tau_{x\Box} ,
\end{equation}
where ${{\mbox{1\hskip-0.22em\relax l}}\,}_\Box$ gives the identity in the
plane $\Box$ and vanishes
outside it, and ${{\mbox{1\hskip-0.22em\relax l}}\,}$ is the usual $SU({N_f})$
identity.  Of course,
this form for $\Sigma$ can deform while
remaining a nontrivial flux tube. In particular, for $\Sigma$ on
$G/H$, the variables $\alpha$ and $\beta$ give coordinates on the
physical plane $R^2$, identified to $S^2$ by the condition $\Sigma
\rightarrow {{\mbox{1\hskip-0.22em\relax l}}\,}$ at spatial infinity ({\em
i.~e. } on the boundary $\alpha \in \{0,
2\pi\}$; $\beta \in\{ 0,\pi\})$. We can deform $\Sigma$ to a radially
symmetric form by identifying $r=\infty$ with this boundary, and
$r=0$ with the center $\alpha = \pi, \beta =
\pi/2$ of the deformation $g$ producing $\Sigma$. This yields
\begin{equation}\begin{array}{c}
\label{Sigralform}
\Sigma(r, \theta) = h(\theta) \ b(r) \ h^{-1}(\theta),
\ \ \mbox{for }\nonumber\\[3pt]
 h(\theta) = \exp{\{i \theta \,\, T_{h\Box}\}},
\ b(r) = \exp{\{i F(r) \,\, T_{b\Box}\}},
\end{array}\end{equation}
with $T_{h\Box}, T_{b\Box}$ as
in our definition \ref{defbh} for $g(\alpha, \beta)$.

We derive this radially symmetric form for $\Sigma$ and fix
$F(r=0)$ as follows. We expand equation \ref{Sigralform} to obtain
\begin{equation}
\label{Sigrexp}
\Sigma(r, \theta) =
{{\mbox{1\hskip-0.22em\relax l}}\,} + \left(\,\,\cos (F/2) - 1 \,\,\right)\
{{\mbox{1\hskip-0.22em\relax l}}\,}_\Box
+ i \sin{(F/2 )} \left(\,\, \cos{\theta}\ \tau_{z\Box} -
\sin{\theta}\ \tau_{x\Box} \,\,\right).
\end{equation}
which can be identified with equation \ref{Sigtpexp}, term by term. This
gives an undeformed relation between the topological description $g(\alpha,
\beta)$ and the spatial form $F(r),\theta$:
\begin{eqnarray}
F(r) &=& 4 \tan^{-1} \left( \,\,[\sin(\alpha/2) \sin\beta]^{-2} - 1\,\,
                 \right)^{-1/2}\nonumber\\
\theta &=& 2 \tan^{-1} \left( \,\cot(\alpha/2)
                \left[ \cos\beta  + \sin\beta
                \left( \,\, [\sin(\alpha/2) \sin\beta]^{-2} - 1\,\,
                 \right)^{1/2}\right]^{-1}\,\right),
\end{eqnarray}
where $\tan^{-1}$ gives values in a single period $[-\pi/2,\pi/2]$.
The arguments to $\tan^{-1}$ range from $0$ to $\pm\infty$ for
$\theta$, and from $0$ to $\infty$ for $F(r)$. Thus $\theta$ covers a
range $[-\pi,\pi]$, suitable for a polar angle, and $F(r)$ ranges from
$0$ to $2\pi$. $F(r)$ assumes the extremes of its range at the
boundary $r=\infty$, where $F(r)=0$, and the center $r=0$, where
$F(r)=2\pi$. These boundary conditions comprise the only relevant
feature of the mapping, the homotopy invariant $F(r=0) - F(r=\infty).$
They establish boundary conditions for any single-winding flux tube of
form \ref{Sigralform}: $F(0) = 2\pi, F(r \rightarrow \infty) = 0$.

Finally, we note that while form \ref{Sigralform} for $\Sigma$ indeed
minimizes gradient energy, potential terms for $\Sigma$ --- like a quark
mass term --- can favor a vev other than $\exp{(i F(r) \,\,
T_{b\Box})}\ $. Potential terms consistent with unbroken $SO({N_f})$
symmetry give a vev we can always diagonalize.  We thus consider
the extending $\Sigma$ so that $F(r) \, T_{b\Box}\ \rightarrow F_d(r)
\,T_d,$ where $F_1 \equiv F$ and $T_d$
varies over the Cartan subalgebra generators \ref{defTd}.
The nonplanar $T_{d>1}$ commute with both $T_{b\Box}$ and $T_{h\Box}$; that is,
they commute with the full embedding $g(\alpha, \beta)$. We may thus
obtain our extension by the simple modification
$
g(\alpha, \beta) \rightarrow g(\alpha, \beta)\ \exp{\{i F_{d>1}(r) \,
T_{d>1}/2\}}.
$
Taking $F_{d>1}(r)$ to vanish as $r \rightarrow \infty$, this does not affect
the behavior of $g$ on the boundary --- hence it leaves the topology
unchanged.  It changes the resulting flux tube only by the desired
overall multiplication $\Sigma \rightarrow \Sigma \ \exp{\{i F_{d>1}(r)\,
T_{d>1}}\}
.$ Thus the true minimum energy flux tube is one of a family of
nontrivial configurations:
\begin{equation}\begin{array}{c}
\label{Sigfinform} \Sigma(r, \theta)=h(\theta) \ b(r) \
h^{-1}(\theta),\ \ \mbox{with }\\[3pt]
h(\theta)= \exp{\{i \theta
\,\, T_{h\Box}\}},\ b(r) =  \exp{\{i F_d(r) T_d\, \}},\\[3pt]
\mbox{and } F_1(r=0) = 2\pi; \ \ F_d(r\rightarrow\infty) = 0.
\end{array}\end{equation}
Which of these candidates is realized remains a question
of dynamics.

\section{Flux Tube Stability and Dynamics}

Studying flux tube dynamics begins with the
question of stability. As minimal model for the Goldstone field
 $\Sigma$, with stable skyrmions, we have
the Skyrme lagrangian
\begin{equation}
\label{SSkyrme}
{\cal L}_0= \frac{F_\pi^2}{16} \,\,{\mbox{tr}\,} \,\partial_\mu \Sigma
\partial^\mu {\Sigma}^\dagger
        +\frac{1}{32e^2} \,\,{\mbox{tr}\,}\,
  {\hbox{\bf [}  {\Sigma }^\dagger \partial_\mu \Sigma,{\Sigma }^\dagger
\partial_\nu \Sigma  \hbox{\bf ]} }^2
\ .
\end{equation}
Unaltered, this lagrangian implies a size instability for all flux tubes.
For under the rescaling $\Sigma(r,\theta) \rightarrow \Sigma (\lambda
r,\theta)$, the tension of a finite flux tube can always decrease.
Specifically, its quadratic contribution stays invariant, while its
quartic rescales by a factor $\lambda^2$ --- leading to an energy
minimized when $\lambda = 0.$ Physically, this corresponds to flux
tubes which diffuse to infinite size to lower their energy.

To stabilize the flux tubes, we must consider modifications of the
minimal forms for $\Sigma$ and ${\cal L}_o$. First, we note that adding
higher derivative gradient terms to ${\cal L}_o$, analogous to the Skyrme
term, cannot both stabilize $\Sigma$ and produce a positive definite
Hamiltonian.  This holds because such terms can be at most second
order in time derivatives, and, by Lorentz symmetry, $r$ and $\theta$
derivatives; thus they can never give tension contributions scaling
more strongly than $\lambda^2$. This leaves only two options for
stabilizing $\Sigma$. First, we can consider a potential for
$\Sigma$, giving a tension component that scales as $\lambda^{-2}$.
Second, we can allow the flux tube to vary along its axis --- giving
$\Sigma$ some $z$-dependence --- to obtain a quadratic contribution to
the energy scaling as $\lambda^{-2}$. The second option, implemented
by exciting zero modes along the flux tube's $z$-axis, has a
structure paralleling that which arises in quantizing the flux tube.
We thus defer a study of its dynamics until the next section
on quantization, and focus on stabilization by a potential. Ultimately
both approaches give similar results: both give as minimal flux tubes
representatives from the family \ref{Sigfinform} with vanishing
$F_{d>1}(r)$. Stabilization by $z$-rotation has a richer
dynamical structure, however --- partly because it introduces
unconstrained rotational parameters into the problem.

A stabilizing potential, on the other hand, arises naturally in our
theory.  By giving the quarks bare Majorana masses, we explicitly
break the $SU({N_f})$ flavor symmetry of our original QCD gauge theory.
This induces a pion mass term,
\begin{equation}
\label{SMass}
{\cal L}_m =       \frac{F_\pi^2 \, m_\pi^2}{16} {\mbox{tr}\,}
        \left(  \Sigma  + {\Sigma}^\dagger - 2\cdot
{{\mbox{1\hskip-0.22em\relax l}}\,} \right)
\ ,
\end{equation}
in the limit of degenerate quark masses. The mass term stabilizes the
flux tube while affecting its dynamics in a simple way, as we discuss
below.

To study dynamics, we calculate the energy density of flux tubes of the
general form \ref{Sigfinform}. The Skyrme action \ref{SSkyrme} gives
a gradient contribution
\begin{equation}
\label{Egradgen}
\rho_{0}= \frac{F_\pi^2}{16}
        {\mbox{tr}\,} \left \{
        \left(\,F'_d\, T_d\, \right)^2
        +\frac{1}{r^2} \,\, \left( \tilde{T}_\Box^2 -
        {\hbox{\bf [} \,F'_d\, T_d\, ,\tilde{T}_\Box  \hbox{\bf ]} }^2 \right)
        \right \} \ ,
\end{equation}
where
\begin{equation}
\label{defTtilde}
{\tilde{T}_\Box} \equiv
b^{-1} (r) \ {\hbox{\bf [} \ T_{h\Box}\ ,\ b(r)\  \hbox{\bf ]} }
\ .
\end{equation}
and $r$ has been rescaled into the dimensionless units $eF_\pi \,
r_{phys}$.  This form for $\rho_0$ follows algebraically, using no
information about the generators $T_d$ and $T_{h\Box}$.
${\tilde{T}_\Box}$ measures the noncommutativity of the radial
generator $ F_d(r)\, T_d$ with its angular counterpart $T_{h\Box}$.
The nonplanar $T_{d>1}$ commute with all generators $T_{h\Box}, T_d$
--- making ${\tilde{T}_\Box}$ and ${\hbox{\bf [} \,F'_d\, T_d\,
,{\tilde{T}_\Box} \hbox{\bf ]} }$
independent of $F_{d>1}$.  This has two consequences: first, nonplanar
terms contribute only the positive definite sum $\sum_{d>1}
{F'_d}^2/2$ to the gradient energy density. Second, the
${\tilde{T}_\Box}$-dependent terms span only a planar $SU(2)$
subgroup, where $T_{d=1} = {{\textstyle\frac{1}{2}}} \tau_{z\Box}$ and
$T_{h\Box} = {{\textstyle\frac{1}{2}}}
\tau_{y\Box}$. They are thus calculable, giving the flux tube gradient
energy density
\begin{equation}
\label{Egrad}
\rho_{0}= \frac{F_\pi^2}{16}
        \left \{ {{\textstyle\frac{1}{2}}}  {\textstyle \sum_{d}} {F'_d}^2
        +\frac{1}{r^2} \left( 1- \cos F_1 \right) \left( 1 + {F_1'}^2 \right)
        \right \} \ .
\end{equation}

The pion mass term \ref{SMass} also contributes to the energy.
Form \ref{Sigfinform} for the flux tube yields its contribution in closed form:
\begin{equation}
\label{EpotN}
\rho_m = \frac{m_\pi^2}{16e^2} 
        \left \{ 2 {N_f} - 4 \cos \left(\sum_{d>1}\omega_d/2\right)
\cos (F_1/2) - 2 \sum_{d>1}\cos\omega_d
        \right \} \ ,
\end{equation}
with
\begin{equation}
\label{defomegad}
\omega_d \equiv \frac{- d \,\, F_d}{\sqrt{2d(d+1)}}
 + \sum_{i>d}\frac{F_i}{\sqrt{2i(i+1)}} \, ,
\end{equation}
where $d = 2, \ldots,{N_f}-1$ and $r$ has again been rescaled to $eF_\pi
\, r_{phys}$.

The minima of this potential fix boundary conditions for
$\Sigma$ at spatial infinity.  Such minima have
$F_{1} = 2\pi m$, consonant with our a priori
boundary condition $F_1(r\rightarrow\infty) = 0$. To minimize with respect to
the nonplanar $\omega_d$, we set
\begin{equation}
\label{omegavac}
\frac{\partial\rho_m}{\partial \omega_d} \sim
 2\sin\left(\sum_{d>1}\omega_d/2\right) \cos (F_1/2)
+ 2 \sin \omega_d = 0\ ,
\end{equation}
with positive second derivatives ${\partial^2 \rho_m} /{\partial
  \omega_d^2}, {\partial^2 \rho_m}/ {\partial F_1^2}$.  This gives the
degenerate family $\omega_d = 2\pi j_d$, for any integers $j_d$
summing to the same parity as $m$. As the simplest representative, we
take $\omega_2 = 2m\pi$ and $\omega_d = 0$ for $d>2$. This
completes our boundary conditions at infinite radius: $F_1 \rightarrow 0$,
$\omega_d \rightarrow 0$, for all $d$; corresponding to $F_d \rightarrow 0$ for
all $d$, in our original variables.

Away from spatial infinity, we must propagate these boundary
conditions inward to obtain flux tubes obeying the full equations of
motion. However, given our boundary conditions $\omega_d(r\rightarrow\infty)
= 0$, these equations trivialize for the nonplanar variables
$F_{d>1}$.  They have independent gradient energy terms \ref{Egrad},
which combine with the potential \ref{EpotN} to give equations of
motion
\begin{equation}
\label{eomnonplan}
\frac{F_\pi^2}{16}\ \frac{1}{r}\,\,\partial_r \left( \, rF'_d \right) =
\frac{\partial \omega_e}{\partial F_d}\
\frac{\partial \rho_m}{\partial \omega_e}
\quad\mbox{for $d>1$.}
\end{equation}
 From equation \ref{omegavac}, the vacuum values $\omega_e =2\pi j_e$
cause the source terms to vanish on the right hand side --- even when
the planar field $F_1$ departs from its vacuum value.  Thus our
boundary conditions, setting $F_d=\omega_d=0$ at infinite radius,
induce only constant solutions as we propagate equation \ref
{eomnonplan} inward to the origin. This gives $F_d$ that vanish
identically for nonplanar $d>1$.

This implies that the minimum energy flux tube for any ${N_f}$ varies
only over the planar $SU(2)$ subgroup. It takes the simple form
\ref{Sigralform}, with $F(r) \equiv F_1(r)$, and has energy density
\begin{equation}
\label{Eflux}
\rho= \frac{F_\pi^2}{16}
        \left \{ {{\textstyle\frac{1}{2}}} {F'} ^2
        +\frac{1}{r^2} \left( 1- \cos F \right) \left( 1 + {F'}^2 \right) +
         \lambda^2 \left( 1 - \cos (F/2)\right) \right \} \ ,
\end{equation}
where $\lambda = 2 m_\pi/e\,F_\pi.$ This determines a nonlinear
equation of motion for $F$:
\begin{eqnarray}
\label{eomplan}
\lefteqn{
\left( \,\, 1 +\frac{4}{r^2} \sin^2 (F/2)\, \right)\
\frac{1}{r}\,\,\partial_r \left( \, rF' \right) =}&&\nonumber\\
&&\sin (F/2)\left\{ \ \frac{2}{r^2}\,\cos(F/2) \
\left( \,\, 1-{F'}^2 \right) \ + \
\frac{8F'}{r^3} \sin(F/2) \ + \ \frac{\lambda^2}{2}
\right\} \ .
\end{eqnarray}

Numerical solutions to equation \ref{eomplan} are shown in Figure 1
for different values of $\lambda$, including the physical
$\lambda_0=0.236$ ($e=2\pi$, $m_{\pi}=138$ Mev and $F_{\pi}=186$ Mev).
Increasing $\lambda$ raises the flux tube's energy density while
shrinking its core size. The asymptotic regimes agree with limits to
equation \ref{eomplan}. Inside the core, $F$ falls linearly from its
boundary value of $2\pi$ at the origin; outside, it scales as the
hyperbolic Bessel function $F \sim x^{-1/2}\,\exp{\{-x\}},$ for $x =
\lambda r/2$. The parameter $2\lambda^{-1}$ thus sets the flux tube's
core size in dimensionless units. In physical units, this gives core
size $m_\pi^{-1}$, sensible for a bound state of Goldstone modes
$\Sigma$.  It carries tension proportional to $F_\pi^2$ --- which is
proportional to the number of colors $N_c$ in a large $N_c$ QCD limit.
This supports its interpretation as a mediator of the confining force
between spinor sources, whose scales as the spinor Casimir $N_c$.
Numerically we find tension $4.6 F_\pi^2$ when $\lambda=\lambda_0$.
As $\lambda$ varies, the tension varies as shown in figure 2.

\section{Quantizing the Flux Tube}

\subsection{ Semiclassical Zero Modes and Quantum Stability}

Under quantization, the flux tube samples not only the ground state
above, but also its zero modes. Recall, from section II, $\Sigma$'s
transformation laws:
$\Sigma \rightarrow a\Sigma a^T$ under
global rotations $a$, giving the residual symmetry $H
= SO({N_f})$ at infinite radius where $\Sigma = {{\mbox{1\hskip-0.22em\relax
l}}\,}$. However, because $\Sigma$
varies nontrivially, global $H$ rotations are not
symmetries within the core. Instead they produce distinct
 degenerate configurations, coincident at spatial infinity. The
zero modes of $\Sigma$ explore these configurations:
\begin{equation}
\label{defSigt}
\Sigma(t, r, \theta) = A(t) \ \Sigma(r,\theta)\ A^{-1} (t) \ ,
\end{equation}
where $A(t) \equiv \exp{\{ieF_\pi\,\omega\, t \, \,n_h T_h\}}$ rotates
in $H$ with dimensionless frequency $\omega$. \footnote{We neglect
  other excitations, such as bending modes.} These zero modes have
rotational energy confined to the string core; an energy that can be
made arbitrarily small, classically, by taking $\omega\rightarrow 0$.

Classically, we calculate this rotational energy from the Skyrme
action \ref{SSkyrme}:
\begin{equation}
\label{Ezms}
\rho_{\omega} = \frac{F_\pi^2 \,\,\omega^2}{16}
{\mbox{tr}\,} \left ( \hat{T}^2 -
        {\hbox{\bf [} \,F'_d\, T_d\, ,\hat{T} \hbox{\bf ]} }^2 -
        \frac{1}{r^2}   {\hbox{\bf [} \,{\tilde{T}_\Box}\, ,\hat{T} \hbox{\bf
]} }^2
        \right) \ .
\end{equation}
Here ${\tilde{T}_\Box},$ from equation \ref{defTtilde},
measures the noncommutativity of radial and angular generators
$F_d(r) T_d$ and $T_{h\Box}$, while
\begin{equation}
\label{defThat}
\hat{T} \equiv b^{-1}(r) \ {\hbox{\bf [} \ h^{-1 } (\theta)\ n_h T_h\ h
  (\theta)\ ,\ b(r)\ \hbox{\bf ]} }
\end{equation}
measures the noncommutativity of $A(t)$ with $\Sigma(r,\theta)$.  We
have retreated to the general form \ref{Sigfinform} --- including
nonplanar contributions --- for $\Sigma(r,\theta)$; thus we retain
complete expressions \ref{Egrad} and \ref{EpotN} for its gradient and
potential energy.  We do this for two reasons: first, we must insure
that the planar vacuum $F_{d>1} =0$ --- favored by the weak pion mass
term --- remains minimal despite quantum fluctuations due to zero
modes; and second, we wish to explore stabilization by $z$-rotation,
where $A = A(z)$.

Calculating the rotational energy \ref{Ezms} is straightforward.  We
expand $2n_h T_h$ as $ n_{(jk)} \,\tau_{y(jk)}$; then note that the
tilde operation, defined in equation \ref{defTtilde}, acts on basis
elements as follows:
\begin{equation}
\label{Ttildebasis}
2 \tilde{T}_{(jk)}  =
\left(  \begin{array}{rr}
&-i{(\pi_{jk}-1)}\\ i(\pi^*_{jk}-1)&
\end{array} \right)_{(jk)} =
( \cos\omega_{jk}-1) \,\,\tau_{y(jk)} +
\sin \omega_{jk}  \,\,\tau_{x(jk)} \ .
\end{equation}
Here $\pi_{jk}$ is given by
\begin{equation}
\label{defomjk}
\pi_{jk} \equiv \exp{\{i\omega_{jk}\}}
\equiv
\exp{\{i F_d (T_{d,\,\,k} - T_{d,\,\,j})\}} \ ,
\end{equation}
where $T_{d,\,\,k}$ denotes the $k$-th diagonal entry in the Cartan
generator $T_d$, from equation \ref{defTd}.
The commutator terms  act on the relevant basis elements $\tau_{x(jk)},
\,\,\tau_{y(jk)}$ to give
\begin{eqnarray}
{\hbox{\bf [} \ T_d\ , \ \tau_{a(jk)}\   \hbox{\bf ]} } &=& -i (T_{d,\,\,k} -
T_{d,\,\,j})\
 \epsilon_{ab}\,\,\tau_{b(jk)} \quad \forall j,k\nonumber\\
{\hbox{\bf [} \ 2\,\tilde{T}_\Box\ , \ \tau_{a\Box'} \   \hbox{\bf ]} } &= &
i p\,(\cos{\omega_\Box} -1 ) \,\,\tau_{a\Box''}
+ i \, \sin{\omega_\Box}\,\,  \epsilon_{ab}\,\,\tau_{b\Box''}
\ , \end{eqnarray}
where $\epsilon_{ab}$ is the two dimensional permutation matrix
$\epsilon_{xy}.$ The second commutator contributes only when
planes $\Box$ and $\Box'$ intersect, with ${\hbox{\bf [} \tau_{y\Box} ,
  \tau_{y\Box'}\hbox{\bf ]} } = ip\,\, \tau_{y\Box''}$ defining $\Box''$ and $p
=
\pm 1$.  Altogether, these expressions imply the rotational energy
\begin{equation}
\label{Erot}
\rho_\omega = \frac{F_\pi^2\omega^2}{16}
\left\{\begin{array}{c}
n^2_{(12)}   \left( 1- \cos F_1 \right) \left( 1 + {F'_1}^2 \right) +
{\displaystyle \sum_{k>j>2}} n^2_{(jk)}
 \left( 1- \cos \omega_{jk} \right) \left( 1 + {\omega'_{jk}}^2
\right) +\\[0.3ex]
{{\textstyle\frac{1}{2}}} {\displaystyle \sum_{k>2}} ( n^2_{(1k)} +  n^2_{(2k)}
)
\left(\begin{array}{r}
 \left( 1- \cos \omega_{1k} \right) \left( 1 + {\omega'_{1k}}^2
        + \frac{1}{2r^2} \left( 1- \cos F_1 \right)  \right) \\ +
 \left( 1- \cos \omega_{2k} \right) \left( 1 + {\omega'_{2k}}^2
        + \frac{1}{2r^2} \left( 1- \cos F_1 \right)  \right)
\end{array}\right)
\end{array}\right\} \ .
\end{equation}

This form for the rotational energy depends on the planar $F_1$, both
explicitly and through the variables $\omega_{jk}$. To extract its
$F_1$ dependence, we note that
\begin{equation}
\omega_{jk} = \omega_{k-1} - \omega_{j-1} ,
\end{equation}
where
\begin{equation}
\omega_0 \equiv  - {{\textstyle\frac{1}{2}}}
\left({\displaystyle\sum_{d>1} \omega_d} \right) \ + \ F_1/2 \ ,\quad
\omega_1 \equiv - {{\textstyle\frac{1}{2}}}
\left({\displaystyle\sum_{d>1} \omega_d}  \right) \ - \ F_1/2 \ .
\end{equation}
and the nonplanar $\omega_{d>1}$ are given by eq.~\ref{defomegad}.
We may thus rewrite the rotational energy as
\begin{equation}
\label{Erotfin}
\rho_\omega = \frac{F_\pi^2\omega^2}{16}
\left\{\begin{array}{c}
n^2_{(12)}   \left( 1- \cos F_1 \right) \left( 1 + {F'_1}^2 \right) +
{\displaystyle \sum_{k>j>2}} n^2_{(jk)}
 \left( 1- \cos \omega_{jk} \right) \left( 1 + {\omega'_{jk}}^2
\right) \\[0.5ex] +
{{\textstyle\frac{1}{2}}} {\displaystyle \sum_{k>2}} ( n^2_{(1k)} +  n^2_{(2k)}
)
\left(\begin{array}{c}
\left( 1 + {S'_{k-1}}^2
        + {{F'_1}^2}/{4} + \frac{1}{2r^2} \left( 1- \cos F_1 \right)
\right) \\
\cdot \left( 1- \cos S_{k-1} \cos(F_1/2)\right) \\
+ F'_1\ S'_{k-1}
\sin S_{k-1}\sin(F_1/2)
\end{array}\right)
\end{array}\right\} \ ,
\end{equation}
where $ S_{k} \equiv \omega_{k} + {{\textstyle\frac{1}{2}}} \sum_{d>1}
\omega_d$.
This depends only on $F_1$ (explicitly) and the nonplanar
$\omega_{d>1}$ (explicitly and through $S_{k-1}$, $\omega_{jk}$).

Neglecting the pion mass term for now, our boundary conditions require
$F_1$ and $\omega_{d>1}$ to approach a minimum of the potential \ref
{Erotfin} at spatial infinity. As in the previous section, these
minima occur when $F_{1} = 2\pi m$ and $\omega_{d>1} = 2\pi
j_d$, for any integers $j_d$ whose sum has the same parity as $m$.  We
take as the simplest representative
$\omega_2 = 2m\pi$ and $\omega_{d>2} = 0$. (The specific case
${N_f} = 3$ has additional degenerate vacua, non-coincident with
those of the pion mass term, given by $\omega_2 = 2m\pi/3$; however,
these reduce to the pion mass vacua above when more families
are added.) Thus our boundary condition $F_1(r\rightarrow\infty) = 0$
induces the full boundary conditions at infinite radius: $F_1 \rightarrow
0$, $\omega_{d>1} \rightarrow 0$; corresponding to $F_d \rightarrow 0$
for all $d$, in our original variables.

The rotational energy $\rho_\omega$ generates spatially-dependent
corrections to the equations of motion for $F_d$. However, these
corrections trivialize for the nonplanar variables $F_{d>1}$, given our
boundary conditions $\omega_d = 0$. We show this as follows: the
rotational energy \ref{Erotfin} introduces an additional term
\begin{equation}
\Delta  = \frac{\partial \omega_e}{\partial F_d}\
\left(\frac{\partial \rho_\omega}{\partial \omega_e}
 - \frac{1}{r}\,\,\partial_r \left(
        \,\, r \,\,\frac{\partial \rho_\omega}{\partial \omega'_e}\right)
\right)
\end{equation}
into the right hand side of the equation of motion \ref{eomnonplan}
for $F_d$. This term is generally quite complicated;
however, when the nonplanar $\omega_{d>1}$ assume their asymptotic
values of zero, it simplifies. For a single $\omega_e$ (before
multiplying by the Jacobian), it becomes
\begin{eqnarray}
\label{eompt}
\Delta_e& = &\omega^2 \left\{
E_{e+1} + {{\textstyle\frac{1}{2}}} {\displaystyle \sum_{k>2}}  E_k\right\}\ ,
\quad
\mbox{with}\nonumber\\
E_k &=&
-  F'_1\ S'_{k-1} \ \sin(F_1/2) - {{\textstyle\frac{1}{r}}}\, \partial_r
\,(\,2r S'_{k-1})\
\left( 1- \cos(F_1/2)\right) \ .
\end{eqnarray}
Asymptotically, where $F_1 =0$, all $\Delta_e$ vanish, and
the full equation of motion \ref {eomnonplan} is again sourceless.
This gives asymptotic solutions for the nonplanar $F_d$ which are
again Bessel functions of order zero --- implying, for finite energy,
that $F_{d>1}$ and all its derivatives (or $\omega_{d>1}$ and all its
derivatives) vanish at infinite radius.

However, given nonplanar terms that obey $\omega_{d>1} = \omega'_{d>1} =
\omega''_{d>1} = 0,$ the perturbations $\Delta_e$
vanish, regardless of the planar field $F_1(r)$. So, as for pion mass
stabilization, equation \ref {eomnonplan} remains sourceless as we
propagate it inward to the origin, giving $F_{d>1}$ that vanish
identically.

Thus the planar vacuum $F_{d>1}=0$ survives quantum fluctuations due
to zero modes. Its static limit retains the form \ref{Sigralform},
with $F(r) \equiv F_1(r)$ varying only over a planar $SU(2)$ subgroup.
It has classical energy, from equation \ref{Erotfin},
\begin{equation}
\label{Ezmfin}
\rho_{tot} = \frac{F_\pi^2}{16}
        \left \{
\begin{array}{c}
         {{\textstyle\frac{1}{2}}} {F'} ^2
        +\left( r^{-2} + \omega^2 n^2_\Box \right)
 \left( 1- \cos F \right) \left( 1 + {F'}^2 \right)
        \ + \lambda^2
        \left( 1 - \cos (F/2)\right)
         \\
+  \omega^2 n^2_{\Box'}\ \left( 1- \cos(F/2)\right)
\left( 1 + {{F'}^2}/{4} + \frac{1}{2r^2} \left( 1- \cos F \right)
\right)
\end{array} \right \}
\end{equation}
where
\begin{equation}
n^2_\Box = n^2_{(12)} \,\, ,\quad  n^2_{\Box'} =
 {\scriptstyle \sum_{k>2}}  n^2_{(1k)} +  n^2_{(2k)}  \ ,
\end{equation}
describe the orientation of the zero mode rotation $A(t)$ relative to
$\Sigma(r, \theta)$.

For a flux tube stabilized by $z$-rotation, where $A = A(z)$, the
rotational terms in $\rho_{tot}$ deform the nonlinear equation of motion for
$F$. Classically, we can rescale $\omega$ to set
$n^2_{\Box} + n^2_{\Box'} = 1$ (a rescaling unnecessary when ${N_f} =
3$).  This gives energy as a monotonic function of $n^2_{\Box}$
--- which measures the component of $\Sigma\,\mbox{'s}$ $z$-dependence
due to slip dislocation, where $\Sigma(z,r,\theta)$ rotates about its
core in a single internal space plane $\Box$, but starts at the
$x$-axis with changing offset angle $\theta_0(z)$.  Its converse,
$n^2_{\Box'}$, measures the twist dislocation in $\Sigma(z,r,\theta)$:
the extent to which $A(z)$ rotates the internal space plane, $\Box
\rightarrow R_{ab}\,(z) \Box$, in which $\Sigma(z,r,\theta)$ cycles about
its core.

Numerical solutions for the $z$-stabilized $F$ are shown in Figures 3
and 4.  Their
asymptotic behavior agrees with analytic limits: inside the core, $F$
falls linearly from its boundary value of $2\pi$ at the origin; while
outside, it scales as the hyperbolic Bessel function $F \sim
x^{-1/2}\,\exp{\{-x\}},$ for $x = \left( \lambda^2 + \omega^2 (1 + 3
n^2_{\Box}) \right)^{1/2} r/2$. That is, the flux tube becomes more
compact as either $\omega$ or $n_\Box$ grows. As in the mass
stabilization case, shrinking core sizes correlate with growing
tensions. Thus we see, from Figure 3, that the value $n_{\Box}=0$ is
favored, and twisting dislocations cost less energy than their
slip-offset counterparts. Figure 4 confirms the flux tube's tendency
both to shrink and to gain energy with increasing rotational frequency, and
shows how significant the rotational deformation of $F(r)$ is. Thus the
 tension not only acquires rotational energy terms,
\begin{equation}
\label{rotten} 
\tau
\rightarrow\ \tau + {{\textstyle\frac{1}{2}}} \omega^2
(\,n^2_{\Box}\,\Lambda_{\Box} + n^2_{\Box'}\,\Lambda_{\Box'}) ) \ ;
\end{equation}
it also contains hidden rotational dependence through deformations of
the ground state tension $\tau$ and moments of inertia
$\Lambda_{\Box}, \Lambda_{\Box'}$. We explore these deformations in
Figure 5, showing that $\tau$ grows linearly with $\omega$, due to
flux tube compression. The moments of inertia instead fall rapidly,
dropping by a factor of five as $\omega$ grows from $0$ to $2$, before
stabilizing at roughly $\Lambda_{\Box} = 10 \ F_\pi^2,
\Lambda_{\Box'} = 4 F_\pi^2$ for $\omega \ge 2\,$ . We show below that
this range $\omega \ge 2$ is typical of quantum rotational excitations
of the flux tube; thus we are justified in neglecting rotational
deformation of $\Lambda_{\Box}, \Lambda_{\Box'}$ over this range.

Finally, we note that the geodesic parametrization assumed
above for the rotation $A(t)$, while useful for discussing the flux
tube's classical limit, does not restrict our analysis. Any function
$A(t)$ over $H= SO({N_f})$ induces the energy \ref{Ezmfin}, with $\omega
\, n_{(jk)}$ defined by
\begin{equation}
\label{defoms}
\omega\, n_{(jk)} \equiv {-i}{(eF_\pi )}^{-1}\,\,
        {\mbox{tr}\,} \ {A}^\dagger \dot{A} \,\, \tau_{y(jk)} \ .
\end{equation}

\subsection{The Quantized Spectrum}

 From the previous section, there are two ways to stabilize the flux
tube: by exploiting the pion mass term in the Lagrangian or by
$z$-rotation.  These give the two classical solutions
$\Sigma(r,\theta)$ (equation \ref{Sigfinform}) and
$\Sigma(r,\theta,z)$ (equation \ref{defSigt}, with $t\rightarrow z$).  Both
have rotational zero modes induced by $A(t)\in H$. We now
quantize the spectrum of these zero modes to find the flux tube's quantum
numbers and lowlying excitations \cite{Balach}.

Quantizing $\Sigma(r,\theta,z)$ is a complicated task, because it
involves two rotations, the $z$ and $t$ rotations. Furthermore,
there is no physical input for the $z$-rotation frequency $\omega$.
Therefore, we restrict ourselves to quantizing the mass-stabilized solution
$\Sigma(r,\theta)$.

Equations \ref{rotten} and \ref{defoms} determine
a two-dimensional Lagrangian for the flux tube:
\begin{equation}
\label{MQSkyrme}
L({\tilde z},t) = -\tilde{\tau}
-{\tilde{\Lambda}_{\Box}\over 2}({\mbox{tr}\,}{A}^\dagger{\dot A}T_{h\Box})^2
 -{\tilde{\Lambda}_{\Box'}\over 2} \sum_{\Box'}\
({\mbox{tr}\,}{A}^\dagger{\dot A}T_{h\Box'})^2\ ,
\end{equation}
where $\tilde{\tau}$ is a ground state tension and
${\tilde z}=eF_{\pi}z$ measures dimensionless length
along the flux tube.
The moments of inertia $\tilde{\Lambda}_{\Box}$ and $\tilde{\Lambda}_{\Box'}$
come from equation \ref{Ezmfin}:
\begin{eqnarray}
\tilde{\Lambda}_{\Box} &=& {\pi\over {e^3F_{\pi}}}
\int\,rdr\ (1+ F^{\prime 2})(1-\cos F)
\nonumber\\
\tilde{\Lambda}_{\Box'} &=& {\pi\over {e^3F_{\pi}}}
\int\,rdr\ \left(1+{F^{\prime 2}/ 4}+ {{\textstyle\frac{1}{2r^2}}}
(1-\cos F) \right)\left(1-\cos(F/ 2)\right)\ .
\end{eqnarray}

We note that form {\ref{MQSkyrme}} for the Lagrangian comes from
only two facts: from the confinement of $\Sigma$
to the planar $SU(2)_{\Box}$ subgroup in internal space;
and from cylindrical symmetry in physical
space. The skyrmion in this theory, which we discuss in the
following section, shares these characteristics.
Thus its one-dimensional Lagrangian $L(t)$ also has form
{\ref{MQSkyrme}}, with $\tilde{\tau}$, $\tilde{\Lambda}_{\Box}$ and
$\tilde{\Lambda}_{\Box'}$ dependent upon skyrmion dynamics.

To quantize such a Lagrangian, we must do two things:
first, we must understand the transformation properties
of $\Sigma$ under the symmetries of $L$;
and second, we must write $L$ in terms of invariants of the
quantized Noether charges of those symmetries.
The first task is facilitated by considering the most
general form of $\Sigma$ on the vacuum manifold:
\begin{eqnarray}
\label{GSigma}
\Sigma(x_i,t)\ &=&\ A(t)\ e^{iF(x_i)n_b(x_i)T_b}\ A^{-1}(t)\nonumber\\
\ \ \ \  &=&\ e^{iF(x_i)n_b(x_i)R_{b'b}T_{b'}}\ ,
\end{eqnarray}
where $R_{b'b}(t)$ is an orthogonal matrix encoding the isospin rotation
due to $A$-conjugation:
\begin{equation}
\label{Rotation}
A^{\dagger}T_iA\ =\ R_{ij}T_j \quad\Rightarrow\quad
R_{ij}(t)\ =\ 2\ {\mbox{tr}\,}(\,\, {{A}^\dagger}(t)\,\, T_{i}\,\, A(t)\,\,
T_j\,\, )\ .
\end{equation}
While $R_{ij}$ is defined for all $SU(N)$ generators $T_i$,
it breaks into block diagonal form, $R_{hh'}\bigoplus R_{bb'}$,
over the space of unbroken and broken generators respectively.

For the general form $\Sigma(x_i,t)$, the Lagrangian
{\ref{MQSkyrme}} involves a general sum
$-{\tilde{\Lambda}_h\over 2}({\mbox{tr}\,}{A}^\dagger {\dot A}T_h)^2$.
This Lagrangian is invariant under global $h\in SO({N_f})$,
implemented by either
left or right multiplications $A\to hA$ or $A\to Ah$.
For infinitesimal transformations
$h_{\epsilon}=1-2i\epsilon_hT_h$, $\Sigma(x_i,t)$
transforms as:
\begin{eqnarray}
\label{DA}
\delta_LA &=& (-i2\epsilon_hT_h)A
\ \ \Longleftrightarrow\ \
\delta_L\Sigma(x_i,t)=-i\epsilon_h\ [\ 2T_h\ ,\
\Sigma(x_i,t)\ ]\nonumber\\
\delta_RA &=& A(-i2\epsilon_hT_h)
\ \ \Longleftrightarrow\ \
\delta_R\Sigma(x_i,t)=-i\epsilon_h\ A\
[\ 2T_h\ ,\ \Sigma(x_i)\ ]\ {{A}^\dagger}\ .
\end{eqnarray}
Therefore the left transformation corresponds to $SO({N_f})$
isospin rotation, and its Noether
charge ${\cal I}_h$ satisfies
an $so({N_f})$ algebra after quantization. The right transformation corresponds
to $SO({N_f})$ isospin rotation in the body fixed frame, with
Noether charge ${\cal I}_h'$.
To understand the physical interpretation of ${\cal I}_h'$,
we calculate the commutator in the second line of equation {\ref{DA}},
obtaining
\begin{equation}
\label{DRA}
A[2T_h,\Sigma(x_i)]{A}^\dagger =
2i f_{hbb'}n_{b}\partial_{n_{b'}}\Sigma(x_i,t)\ .
\end{equation}

For the specific cases of flux tube and skyrmion, $n_b (x_i)$ has
two special properties: $n_b = n_{b\Box}$, lying only in an $SU(2)$
subplane of $SU({N_f})$; and $n_b (x_i)$ is spatially axisymmetric,
depending linearly on spatial direction components $u_x$ and $u_y$.
These properties
identify planar isospin rotations with spatial
$z$-rotations. We see this in two steps: the planarity of $n_b$
reduces equations \ref{DA} and \ref{DRA}, when $h = h\Box$, to
\begin{equation}
\delta_R \,\, \Sigma (r, n_b(r, u_i) )
= -2\epsilon \,\, f_{h\Box\,b\Box\,b'\Box}\,\,
n_{b\Box} \,\, \partial_{n_{b'\Box}} \ \Sigma (r, n_b'(r, u_i) ) \ ,
\end{equation}
which the Jacobian relating $n_b$ to $u_i$,
at fixed $u_z$, equates with spatial $z$-rotation :
\begin{equation}
\delta_R \,\, \Sigma (r,  u_i)
= -2\epsilon \,\, \epsilon_{zij}\,\,
u_i \,\, \partial_{u_{j}} \ \Sigma (r, u_i)  \ .
\end{equation}
Thus ${\cal I}'_{\Box}$, which generates planar body-centered isospin
rotations $R$, must equal the defect's spin operator $J_z$, generating
spatial $z$-rotations. Other isospin rotations, which destroy the
planarity of $n_b$, are distinct from other spatial rotations, which
change the defect's axis of axisymmetry.

Under right and left infinitesimal transformations of $A(t)$, the
rotations $R_{b'b}$, which give physical coordinates for $\Sigma(x_i,t)$
on the vacuum manifold, transform as
\begin{eqnarray}
\label{LRR}
\delta_LR_{b'b}\ &=&\ -2\epsilon_hf_{hb'c}R_{cb}\nonumber\\
\delta_RR_{b'b}\ &=&\ -2\epsilon_hR_{b'c}f_{hcb}\ .
\end{eqnarray}
That is, $R_{b'b}$ is left or right multiplied
by ${({\tilde h}_{\epsilon})}_{bc}=1+i\epsilon_h(2if_{hbc})$.
This shows how isospin rotation acting directly on the space
of broken generators
can obey an $so({N_f})$ algebra --- specifically the $so({N_f})$
algebra which occurs as a subalgebra
of $su({N_f})$ in the adjoint representation,
restricted to the block-diagonal component $2if_{hbc}$.

To find ${\cal I}_h$ and ${\cal I}_h'$,
we compute the variation of the Lagrangian under
the two infinitesimal transformations:
\begin{equation}
\label{DL}
\delta_LL = {\dot{\epsilon_h}}{\cal I}_h\ \ \ \ \ \ \
{\hbox{and}}\ \ \ \ \ \
\delta_RL = {\dot{\epsilon_h}}{\cal I}_h'\ .
\end{equation}
This gives
\begin{equation}
\label{LR}
{\cal I}_h' = i\tilde{\Lambda}_h {\mbox{tr}\,}({A}^\dagger {\dot A}T_h)\ \
{\hbox{(no sum),}}\
\ \ \ {\hbox{and}}\ \ {\cal I}_h=R_{hh'}{\cal I}_{h'}'\ .
\end{equation}

We can now write the defect Hamiltonian in terms of the physical
quantized Noether charges: the isospin ${\cal I}$ and angular
momentum ${\cal J}$.  Combining the Lagrangian {\ref{MQSkyrme}} with
Noether charges {\ref{LR}} yields a Hamiltonian dependent on ${\cal
I}_{\Box}'^2$ and ${\cal I}_{\Box'}'^2$, where $\Box'$ are planes
intersecting $\Box$ in only one line. However, the defect
$\Sigma(r, u_i )$ is invariant under a subgroup
$SO({N_f}-2)$ of $SO({N_f})$, acting on the subspace $\Box_{\bot}$
orthogonal to $\Box$.  This implies two things: ${\cal I}'^2 =
{\cal I}_{\Box}'^2 + \sum_{\Box'} {\cal I}_{\Box'}'^2$; and
the defect's allowed quantum states ${\cal I}'$ must contain a singlet
under the $SO({N_f}-2)$ subgroup. For ${N_f} > 3$, this excludes spinor
representations --- constraining the skyrmions and flux tubes to
be bosonic.  Lastly, because ${\cal I}$ is simply a rotation of ${\cal
I}'$, and ${\cal I}_{\Box}'= {\cal J}_z$, we write the Hamiltonian as
\begin{equation}
\label{finalH}
H (\tilde{z}, t) =\tilde{\tau}+
{1\over{2{\tilde\Lambda}_{\Box}}} {\cal J}_{z}^2
+{1\over {2{\tilde\Lambda}_{\Box'}}} (\ {\cal I}^2 - {\cal J}_{z}^2\ ) \ ,
\end{equation}
subject to the constraints above: ${\cal I}_{\Box}'= {\cal J}_z$,
with bosonic defects for ${N_f} > 3$.

Using the numerical results of Section IVA, we may be more explicit
about the flux tube's quantized energy levels. Relating the rotational
energies of equation \ref{finalH} to their counterparts in the tension
\ref{rotten} (for dimensionful $z$) gives
\begin{equation}
\omega^2 = (eF_\pi)^4 \
( \,\, {{\cal J}_z}^2/4{\Lambda_{\Box}}^2 + ( \,\, {\cal I}^2 -
{{\cal J}_z}^2\,\, )/4{\Lambda_{\Box'}}^2\,\, ) \approx 4{{\cal
J}_z}^2 +25 ( \,\, {\cal I}^2 - {{\cal J}_z}^2\,\, )\ .
\end{equation}
Thus $\omega$ assumes values $\omega \ge 2$ for quantized zero modes.
Over this range, $\tilde{\Lambda}_{\Box} $ and
$\tilde{\Lambda}_{\Box'}$ vary little but $\tilde{\tau}$ is described
by the linear fit $\tilde{\tau} = (9.0 + 3.7 (\omega - 2))\,\,
F_\pi/e$, as shown in Figure 5.  Thus our model predicts
\begin{equation}
H (\tilde{z}, t) = \frac{F_\pi}{2\pi} \ \left (1.6 + 3.7\,\, \sqrt{
4{{\cal J}_z}^2 +25 ( \,\, {\cal I}^2 - {{\cal J}_z}^2\,\, )} +
20 {{\cal J}_z}^2 + 50 ( \,\, {\cal I}^2 - {{\cal J}_z}^2\,\, )
\right )
\end{equation}
for the allowed excited states $({\cal I}, {\cal J}_z)$, and
\begin{equation}
H (\tilde{z}, t) = 4.6 \ \frac{F_\pi}{2\pi}
\end{equation}
for the ground state of the flux tube.

\section{Flux Tube Interactions}

We now consider how flux tubes interact with fundamental objects in
our low energy theory. First, we establish that flux tubes can be
Alice strings --- forcing some charges and particle wave functions to
become double-valued. Like gauge theories with Alice strings and
monopoles, our sigma model has twisted flux loops which form point
defects.  We construct these defects --- the skyrmions --- and
demonstrate their nontrivial topology. We then consider how they
interact with the flux tubes, showing that only topologically trivial
combinations of two flux tubes can end on skyrmions. This suggests the
physical interpretation that, while the baryons in this theory are not
confined, the spinor sources which combine to form them are, with
confinement mediated by the $Z_2$ flux tubes joining them.

First we consider twisted flux tubes. We note that our flux tubes may
share the defining property of Alice strings, which arise in the
symmetry-breaking of certain gauge theories. \cite{Alice} Alice
strings have unbroken symmetries which preserve a local vev but cannot
be extended globally, since they become multivalued when parallel
transported around the spatially varying vev of the string.  This
``Alice'' nature is not topologically invariant, but depends on the
specific Wilson line integral producing the string's asymptotic
winding.  Whether our flux tubes have this trait is similarly
ambiguous: using the physical embedding \ref{Sigralform} of the flux
tube, we can take $g(r,\theta) = h(\theta)\,\, b(F(r)/2)$. This gives
parallel transported unbroken generators $g\,\, T_{h\Box'} \,\,
g^{-1}$ that are double-valued in $\theta$ for each radius.  However,
all generators can be made single-valued by choosing instead
$g(r,\theta) = h(\theta)\,\, b(F(r)/2)\,\, h(\theta),$ which induces the same
flux tube $\Sigma$.

Because $\Sigma$ can be viewed as an Alice string, we might expect it
to share a property of gauged Alice strings in models with monopoles:
that twisted string loops support monopole charge. For gauged Alice
strings, this is necessary, as monopoles alter their charge when
passing through string loops; moreover, specific examples indicate
that twisted Alice loops can carry a single unit of monopole charge.
\cite{Almon} For global Alice strings, the suggestion arises by
analogy.  In our case, it gains further support from the hidden gauge
character of the low energy global theory. Specifically, we can recast
the flavor-dependent quark mass $\Sigma$ in the usual way, as an
interaction between flavor gauge fields $g^{-1} (r,\theta)
\,\,\partial_\mu \,\, g(r,\theta)$ and shifted quarks $Q_{L} =
g(r,\theta) \,\, q_{L}.$ \cite{Balach} This accomplishes two goals: it
equates skyrmion number with anomalous baryon current, while
identifying the flux tube with a gauged string, whose Wilson
line integral is $ {\rm P}\exp{ \left ( \oint\vec{A} \cdot d\vec{l}\
\right ) }.$ However, the Alice ambiguity persists: choosing $g(r,\theta)$
double-valued gives a gauge theory with Wilson loop $U(2\pi)= {
  {\mbox{1\hskip-0.22em\relax l}}\,} - 2 \,\,{
  {\mbox{1\hskip-0.22em\relax l}}\,}_\Box$, making both the quarks
$Q_{L}$ and the generators $T_{h\Box'}$ double-valued.  A
single-valued choice for $g(r,\theta)$ produces instead single-valued
quarks and generators, with Wilson loop $U(2\pi)= {
  {\mbox{1\hskip-0.22em\relax l}}\,}$.

This ambiguity can be resolved physically, by considering adiabatic
transport of quarks around the flux tube. \cite{wilczek} Under such
transport, the quarks remain in their mass eigenstates. These are
governed by two terms: a flavor-independent bare Majorana mass $M$
(breaking $SU(N_f) \rightarrow SO(N_f)$ explicitly) and the
flavor-dependent Majorana mass $\mu\Sigma$, where $\mu$ is the
$SU(N_f)$-breaking vev. The two determine mass eigenstates that are
double-valued in $\theta$. However, the states have mass splitting
$\Delta m^2 = -2i M (\mu - \mu^*) \sin(F/2)$, where we have chosen $M$
real. Thus the quarks are degenerate, and unaffected by transport
around the string, unless the vev $\mu$ is misaligned in phase with
the bare mass $M$. In that case, quarks at finite radius have
double-valued wave functions and Aharonov-Bohm scatter off the flux
tube, which we identify as an Alice string. \footnote{We consider here
  quarks whose trajectories remain well-separated from degeneracies at
  the origin; that is, for which $r^2 >> m_i^{-2},$ where $m_i$ are
  the quark mass eigenvalues. Thus we rely on an asymptotic regime
  inside the flux tube, in which $m_i^{-1} << m_\pi^{-1}.$ The flux
  tube background affects general quark wave functions more subtly.
  Since the mass eigenstates depend explicitly on $\theta$,
  off-diagonal $\hat{\theta} \cdot \hat{\tau}$ terms appear in the
  quark equations of motion, mixing mass eigenstates at small radius.
  This causes corrections to maximal Aharonov-Bohm scattering, much as
  in \cite{wilczek}.}

Motivated by its gauge theoretic counterpart, we calculate the
skyrmion number of a twisted flux tube:
\begin{equation}
\Sigma(z, r, \theta) = A(z) \ \Sigma(r,\theta)\ A^{-1} (z) \ ,
\end{equation}
where $A(z) = \exp{(i z\, l\, n_h T_h)} \in H.$ Imposing
$2\pi$-periodicity in $z$ constrains $l$: for planar $n_h=n_\Box$, $l$
must be integral; for nonplanar $n_h$, it must be even. Thus any
twisted flux loop can deform to the planar flux loop $ \Sigma(z, r,
\theta) = h(\theta + lz) \ b(F(r)) \ h^{-1} (\theta + lz)$, which
takes values only in the two-flavor subspace $SU(2)/SO(2) \sim S^2.$
This simplification allows us to identify the loop's $\pi_3$ index
with its Hopf number, {\em i.e.} the linking number between any
two fibers of constant $\Sigma$ in physical space. As discussed in
\cite{wuetc}, this linking number is precisely $l$ --- the number of
times a nontrivial fiber $\Sigma_0$ twists around the loop's core,
which has $\Sigma= -{ {\mbox{1\hskip-0.22em\relax l}}\,}_\Box$. Thus
flux loops with an $l=1$ planar twist form fundamental skyrmions; flux
loops with nonplanar twist have $l>1$, which is trivial for $N_f > 3$.

A nicer parametrization of the skyrmion stems from the exact sequence
\[
\pi_3\left( SO(N_f)\right)  \ \  \rightarrow \ \
\pi_3\left( SU(N_f)\right)  \ \  \rightarrow \ \
\pi_3\left( SU(N_f)/SO(N_f)\right)\ \  \rightarrow \ \
\pi_2\left( SO(N_f)\right) = 0 \  .
\]
This implies that fundamental skyrmions in the theory can be
constructed from the fundamental skyrmions
\begin{equation}
g(r, \hat{u} ) = \exp{(iF_s (r) \ \hat{u}_i\,\, T_{i \Box} )}
\end{equation}
in $SU({N_f})$. Here $r$ and $\hat{u}$ are the radius and unit direction
vector in 3-space, and $F_s (r)$ approaches
$2\pi$ at $r=0$ and zero at $r=\infty$. This determines an
axisymmetric skyrmion
\begin{eqnarray}
\label{skyrmion}
\lefteqn{\Sigma_s = {{\mbox{1\hskip-0.22em\relax l}}\,} - 2 \sin^2 (F_s/2)\,\,
(1 - u_z^2) \,\,{{\mbox{1\hskip-0.22em\relax l}}\,}_\Box
           + 2i\, \sin (F_s/2) \,\, \cdot } \nonumber\\
&& \quad\left(
\cos(F_s/2) \left(u_y \ \tau_{x\Box} + u_x \ \tau_{z\Box} \right)
+ \sin(F_s/2) \,\,u_z
\left(-u_x \ \tau_{x\Box} + u_y \ \tau_{z\Box} \right)
 \right)\ ,
\end{eqnarray}
after a global spatial rotation fixing the
$z$-axis as the axis of spatial axisymmetry. In the $xy$-plane, this
gives
\begin{equation}
\Sigma_s(z=0) = {{\mbox{1\hskip-0.22em\relax l}}\,} + (\cos F_s - 1) \
{{\mbox{1\hskip-0.22em\relax l}}\,}_\Box
           + i \,\sin F_s\  \left(
 u_y \ \tau_{x\Box} +
 u_x \ \tau_{z\Box} \right).
\end{equation}
Comparing with equation \ref{Sigrexp} for the flux tube allows us
to identify the angular winding $n_b (\theta)\,\, T_{b \Box}$ of a flux
tube with that of a skyrmion in the $xy$-plane. Thus, if their
radial boundary conditions coincided, we could deform the lower
hemisphere of the skyrmion into a flux tube.  However, because
both $F$ and $F_s$ vary from $0$ to $2\pi$ over this plane, the
skyrmion cannot end in a single flux tube. The boundary
conditions instead allow the skyrmion to join only to flux tube
configurations where $F(r)$ ranges from $0$ to $4\pi$ --- that is,
configurations with two flux tubes, deformable to the trivial
configuration. Thus skyrmions cannot be confined in this theory.

However, objects which combine to form skyrmions can
interact with the flux tubes. Such ``half-skyrmions'' could arise as
external spinor sources in the underlying theory. They should
be confined, as fundamentals cannot screen them. As mappings on $G/H$,
they appear in our theory precisely as half-skyrmions, that is, as
objects of the form \ref{skyrmion} with $F_s(r)$ ranging from $0$ to
$\pi$. Such objects are not defects in the conventional sense, since
they have linearly divergent energy --- just like an unscreened point
source.  Their boundary conditions allow them to join to their
opposite winding counterparts via single flux tubes, confining their
linearly divergent energy to a length scale set by the tube
length.  We thus see that confinement of sources in an
$SO(N_c)$ gauge theory can induce a relic phenomenology, which persists
in the low energy Skyrme model.

\acknowledgements
 We thank Aneesh Manohar, David Kaplan, Tom Imbo, Sidney
Coleman and Glenn Boyd for helpful conversations.  This work was
supported by the Department of Energy under Grant
No.~DOE-FG03-90ER40546 and by the University of California
President's Postdoctoral Fellowship program.

\appendix
\section*{}

In this appendix, we show that trivializations \ref{gformgen} of the
loop $h^2(\alpha)$ produce minimal flux tubes only when
$\tilde{h}(\beta) = {{\mbox{1\hskip-0.22em\relax l}}\,}$.  We do this by
generalizing the analysis
following equation \ref{gformgeod} for arbitrary $\tilde{h}(\beta)$
--- only to find that consistency and energy considerations restore
the restriction $\tilde{h}(\beta) = {{\mbox{1\hskip-0.22em\relax l}}\,}$.

Note that conjugation by $\tilde{h}(\beta)$ rotates
$2T_h$ in $so({N_f})$:
\begin{equation}
\label{hconj}
 n_{(jk)}(\beta) \,\,  \tau_{y(jk)}\rightarrow
\,\, u_{(jk)}(\beta) \,\,  \tau_{y(jk)}, \quad \mbox{where}
\,\, u_{(jk)}(\beta) \,\, \equiv\tilde{h}_{(jp)} (\beta) \,\, n_{(pq)}
\,\,\tilde{h}_{(kq)}(\beta).
\end{equation}
Since $\tilde{h}(\beta)$ is real and orthogonal, $u_{(jk)}(\beta)$ is
just a real unit vector, like $n_{(jk)}.$ $b(\beta)$ then conjugates
this rotated generator, giving Equation \ref{rotTh} with the
substitution $n_{(jk)} \rightarrow u_{(jk)}(\beta). $ This fully conjugated
generator must become $-2T_h$ at $\beta=\pi$, to produce the loop
$g(\alpha,\beta=\pi) = h^2(\alpha)$. This occurs only if $u_{(jk)}
(\pi) = \lambda_{(jk)} n_{(jk)},$ for each $(jk)$, with
\begin{equation}
\label{deflamjk}
\lambda_{(jk)} =
\left\{
\begin{array}{ll}
- \exp{(i l\pi)} \,\, & \mbox{for $(jk) = (12)$}\\
- \exp{(i l\pi/2)} \,\, & \mbox{for $j\le2, k\ge 3$}\\
- 1 & \mbox{for $2<j<k$}.
\end{array}
\right.
\end{equation}
We note that these matching conditions depend only on a basis choice
fixing $b(\beta)$ in the $(12)$ plane. Rotations of $b(\beta)$ within
that plane, for example taking $T_b \rightarrow {{\textstyle\frac{1}{2}}}
\tau_{x(12)}$,
change the details of equation \ref{rotTh}, but not the coefficients
of $\tau_{y(jk)}$ and its orthogonal generators; and these
coefficients alone fix the matching \ref{deflamjk}.

These matching conditions cannot be achieved by orthogonal
conjugation of an arbitrary generator $T_h$. A first constraint stems
from the reality of $\lambda_{(jk)}$. This allows a nonvanishing
$n_{(jk)}$ only when the relevant exponential gives $\pm 1$. For
$l=1$, this forces $n_{(1k)} = n_{(2k)}=0$ for $k>2$ --- splitting
both $h(\alpha)$ and $\tilde{h}(\beta) \,\,h(\alpha) \,\, \tilde{h}
^{-1}(\beta) $ into two commuting blocks. The $2<j,k$ block is
unaffected by $b$-conjugation, and changes $g$ only by an overall
right multiplication, which leaves $\Sigma = g g^T$ invariant.  Thus, for
$l=1$, $T_h ={{\textstyle\frac{1}{2}}}\tau_{y(12)}$ again produces all
nontrivial loops $g$.

Equation \ref{hconj} places more subtle constraints on $T_h$,
related to consistency of the diagonal reduction $u_{(jk)} (\pi) =
\lambda_{(jk)}n_{(jk)},$ for all $(jk).$ This is solved, for
independent $n_{jk}$, only when $\tilde{h}_{(jp)} (\pi)$ is itself
diagonal: $\tilde{h}_{(jp)} (\pi) =\lambda_j\delta_{(jp)},$ with
$\lambda_j$ real. This constrains the matching conditions attainable
by nonzero $n_{(jk)}$ to $\lambda_{jk} = \lambda_j\lambda_k .$ Note
that our $n_{jk}$ are effectively independent, since the relation
\ref{deflamjk} holds for all global rotations of $n_{(jk)}$ that leave
$b(\beta)$ in the $(12)-$plane. Such rotations map a single generator
in any of the three matching classes --- $(jk)$ coincident with
$(12)$;$(jk)$ intersecting $(12)$ in a single line; and $(jk)$
disjoint from $(12)$ --- to the entire class, in arbitrary linear
combinations. Thus, for $l=2,$ only two possibilities are consistent
with the matching conditions \ref{deflamjk}.  We can have nonvanishing
$n_{(jk)}$ in the cross planes $(1k),(2k),$ with $k\ge 3.$ However,
this precludes nonvanishing $n_{(jk)}$ outside the cross planes:
$\lambda_{(1k)} = \lambda_{(2k)} = 1$, for all $k\ge 3$, implies
$\lambda_{(jk)}=1$ for all $j,k$.  Thus we must set $n_{(jk)}=0$
outside cross planes to obey \ref{deflamjk}.  Similarly, nonzero
$n_{(12)}$, with $\lambda_{(12)} = -1$, implies $\lambda_{(1k)} =
-\lambda_{(2k)}$ for $k\ge 3$ --- forcing $n_{(jk)}=0$ in the cross
planes to obey \ref{deflamjk}. The case $l=2$ thus produces no new
candidates for $T_h$: either $n_{(jk)}$ vanishes outside the planes
$(jk)$ which intersect $(12)$ in a single line, or $n_{(jk)} =
e_{(12)}$ in the $(12)$ plane (discarding a right multiplication $g
\rightarrow gh'$ as in the $l=1$ case above).

The only surviving consequence of $\tilde{h}-$conjugation, then, is
an overall rotation $u_{(jk)}(\pi) = \pm n_{(jk)}.$ For the positive
sign, $\tilde{h}(\beta)$ commutes with $h^{-1}(\alpha)$ at both
endpoints in $\beta$, so choosing $\hbox{\bf [}\tilde{h}(\beta),
h^{-1}(\alpha)\hbox{\bf ]} = 0$ for all $\beta$ clearly minimizes gradient
energy.  For
the negative sign, a similar reduction occurs for $b(\beta)$. For then
$b(\beta)$ commutes with $\tilde{h}(\beta) \,\, h^{-1}(\alpha)
\,\,\tilde{h}^{-1}(\beta) $ at both endpoints, and we can deform $g$ to
lie entirely in $H$. Thus only the case $\tilde{h}(\beta) =
{{\mbox{1\hskip-0.22em\relax l}}\,}$
induces a nontrivial defect $\Sigma \in G/H$ of minimal energy.

\begin{figure}
\caption{
  a.) Flux tube solutions $F(r)$ for $\lambda$ values $0.2,0.236,1.0,$
  and $2.0$, with mass stabilization only ($\omega=0$).  The dotted
  line corresponds to the physical $\lambda_0=0.236$. Note that
  core size shrinks with increasing $\lambda$. \quad b.) The above
  flux tubes' energy density $\rho$. Note
  that tension grows with the coupling $\lambda$.  }
\label{one}
\end{figure}

\begin{figure}
\caption{
  Dependence on stabilizing mass ($\omega=0$). Solid line shows
  variation of the total tension $\tau = \int\,\,d^2 r\,\,\rho$ as a
  function of $\lambda$.  Dotted line shows the gradient contribution
  $\tau_0$ only. The physical $\lambda_0=0.236$ is marked.  }
\label{two}
\end{figure}

\begin{figure}
\caption{
  Zero mode dependence on rotational direction.
  Solutions for $\lambda=\lambda_0,$ with fixed frequency $\omega = 1$,
  and varying planar component $n_\Box$.  Note that core size
  shrinks and tension grows as the planar component $n_{\Box}$
  increases from $0$ to $1$. This implies that twisting dislocations
  are favored over slipping ones (see text). Allowing $\lambda$ to vanish
  has negligible effect on $F$ and $\rho_{tot}$.}
\label{three}
\end{figure}

\begin{figure}
\caption{
  The rotating mass-stabilized flux tube ($\lambda=0.236$). For
  $n_\Box=0$, shows the flux tube's deformation due to rotational zero
  modes.  As for other values of $\lambda $, the flux tube's core size
  shrinks and tension grows as the rotation frequency $\omega$
  increases from $0.5$ to $1.5$. Rotation with a nonzero planar
  component $n_\Box$ displays the same trend, but begins with more
  compact and energetic flux tubes when $\omega= 0.5$ (cf. Fig.~3).  }
\label{four}
\end{figure}

\begin{figure}
\caption{
  Rotational deformation, for $n_{\Box}=0.5$. Dotted line shows
  $\lambda_0=0.236$; solid lines show the $\lambda=0$ and
  $\lambda=1$ cases.  a) Variation of the ground state tension $\tau$
  with rotational frequency $\omega$. (Note that $\tilde{\tau}$, the
  tension conjugate to $\tilde{z}$ of Section IV, is given by
  $(eF_\pi)^{-1} \tau$).  \quad b) and c) Variation of the planar and
  nonplanar moments of inertia $\Lambda_{\Box}$ and $\Lambda_{\Box'}$.
  (Note that $\tilde{\Lambda}_{\Box}$ and $\tilde{\Lambda}_{\Box'}$ of
  section IV are given by $4 (e F_\pi)^{-3}$ times their counterparts
  plotted here.)  }
\label{five}
\end{figure}


\begin{references}

\bibitem{wittenon}
E.~Witten, Nucl.\
Phys.\ {\bf B223}, 433 (1983).

\bibitem{shortskyrme}
K.~Benson, A.~Manohar, and M.~Saadi, submitted to Phys.\ Rev.\ Lett.


\bibitem{Balach}
A pedagogical introduction to the conventional Skyrme model (with
Dirac fermions) appears in A.~P.~Balachandran, Proceedings of the Yale
Theoretical Advanced Study Institute ``High Energy Physics 1985'',
ed.~M.~Bowick and F.~G\"ursey, World Scientific, Singapore, 1. We
approach the issues of quantization and anomalous baryon current
in a similar spirit.

\bibitem{Alice}
M.~Alford, K.~Benson, S.~Coleman, J.~March-Russell and F.~Wilczek,
Phys.\ Rev.\ Lett.\ {\bf 64}, 1632 (1990), Nucl.\ Phys.\ {\bf B349},
414 (1991); J.~Preskill and L.~Krauss, Nucl.\ Phys.\ {\bf B341}, 50
(1990); A.~S.~Schwarz, Nucl.\ Phys.\ {\bf B208} 141 (1982);
A.~Balachandran, F.~Lizzi, V.~Rodgers, Phys.\ Rev.\ Lett.\ {\bf 52}
 1818 (1984).

\bibitem{Almon}
M.~Bucher, H.~K.~Lo, and J.~Preskill, Nucl.\ Phys.\ {\bf B386} 3 (1992)
; M.~Alford, S.~Coleman, J.~March-Russell and F.~Wilczek,
unpublished (1991).

\bibitem{wilczek}
J.~March-Russell, et.~al.,
Phys.\ Rev.\ Lett.\ {\bf 68}, 2567 (1992).

\bibitem{wuetc}
  Y.~Wu and A.~Zee, Nucl.\ Phys.\ {\bf B324} 623 (1989); F.~Wilczek
  and A.~Zee, Phys.\ Rev.\ Lett.\ {\bf 51}, 2250 (1983).


\end{references}
\end{document}